\documentclass[12pt]{article}
\usepackage{amssymb,amsmath,epsfig}
\allowdisplaybreaks
\begin{document}
\title{\bf Axial Dissipative Dust as a Source of Gravitational Radiation in $f(R)$ Gravity}
\author{M. Sharif \thanks{msharif.math@pu.edu.pk} and Aisha Siddiqa
\thanks{aisha.siddiqa17@yahoo.com}\\
Department of Mathematics, University of the Punjab,\\
Quaid-e-Azam Campus, Lahore-54590, Pakistan.}

\date{}

\maketitle
\begin{abstract}
In this paper, we explore the source of gravitational radiation in
the context of $f(R)$ gravity by considering axially symmetric
dissipative dust under geodesic condition. We evaluate scalars
associated with electric and magnetic parts of the Weyl tensor for
both non-spinning (at the center) and spinning (in the surrounding
of the center) fluids of the configuration. For this purpose, we use
the evolution as well as constraint equations for kinematical
quantities and Weyl tensor. Finally, we investigate the existence of
gravitational radiation through super-Poynting vector. It is found
that the fluid is not gravitationally radiating in the non-spinning
case but it is gravitationally radiating for the spinning case.
\end{abstract}
{\bf Keywords:} $f(R)$ gravity, Axial source, Gravitational
radiation.\\
{\bf PACS:} 04.50.Kd; 04.40.Nr; 04.40.Dg.

\section{Introduction}

Gravitational waves (GWs) are ripples in the curvature of spacetime
produced by moving objects and gravitational radiation is the energy
carried by these waves. However, gravity is the weakest of the
fundamental forces so these waves are not detectable even through
sensitive detectors unless produced by the massive stellar objects.
One of the important features of GWs is that if we could detect and
observe these waves then it will lead new observational techniques
for different astrophysical phenomena. Recently, LIGO scientific
collaboration and Virgo collaboration \cite{1} provided the first
observational evidence for GWs originating from a pair of merging
black holes. Hawking \cite{2} showed that there is an upper bound
for the energy of gravitational radiation emitted by the collision
of two black holes. Wagoner \cite{3} investigated the gravitational
radiation produced by accreting neutron stars. Flanagan and Hughes
\cite{4} provided a comprehensive study of many important aspects
about theory of GWs. They found that linearized theory is
appropriate to explain GWs propagation, interaction of these waves
with detectors and emission of GWs from a source. They also
discussed different formalisms to deal with the situations where
linearized treatment is not sufficient.

Dust solutions are considered to be ideal models for astrophysical
objects in which particles are assumed to interact only
gravitationally. Ellis \cite{34} investigated dynamics of dust fluid
and found some new solutions of the field equations. Lemos \cite{35}
showed that naked singular solutions formed by gravitational
collapse of radiation and dust are same in nature. Dissipation is an
important phenomenon in the evolution of stellar objects. It is, in
fact, the only mechanism that leads to a star away from hydrostatic
equilibrium by neutrino emission. Although, pressure is present in a
physical process with dissipative fluid, however, some authors
neglect it to see the effects of dissipation on the system as well
as for analytical solution. The dissipative dust cloud has been
considered by many researchers \cite{36} to investigate the causes
of energy density inhomogeneity in GR as well as modified gravity
theories. Herrera \textit{et al.} \cite{14} discussed the existence
of gravitational radiation in dissipative dust. Thus it would be
worthwhile to study the evolution of celestial bodies through
dissipative dust.

It is well-known that astrophysical systems can be rotating and
hence their exterior solutions cannot be exactly spherically
symmetric. The region outside a rotating stellar body can be
represented only by an axial symmetric spacetime, the most
compatible for the interior region \cite{6}. According to Bhirkoff
theorem, spherically symmetric spacetime is not appropriate while
axially symmetric spacetime is compatible with the existence of
gravitational radiation \cite{5}. Herrera {\it et al.} \cite{14,13}
investigated the presence of gravitational radiation in general
relativity (GR) for perfect and dissipative dust fluid with axial
symmetry under the geodesic condition. They showed that both types
of fluid do not produce gravitational radiation.

The Weyl tensor represents that part of curvature which cannot be
determined by matter locally. However, the contracted Bianchi
identities serve as field equations for the Weyl tensor and give the
part of curvature depending on matter \cite{6}. This approach is
used to discuss gravitational radiation in the literature \cite{7}.
Matte \cite{9} proposed electric and magnetic parts of the Weyl
tensor to show an analogy between gravitational waves and
electromagnetic waves. Bertschinger and Hamilton \cite{10} derived
the evolution equations for these parts of the Weyl tensor. Maartens
and Basset \cite{11} introduced super-Poynting vector to describe
the flux of super energy which depends upon the magnetic and
electric parts of the Weyl tensor. When super-Poynting vector is
non-zero it is defined to be a state of intrinsic gravitational
radiation \cite{12}.

The mystery of dark energy leads to the concept of modified theories
of gravity. $f(R)$ gravity is one of the viable modified theories in
which Ricci scalar in the Einstein-Hilbert action is replaced by its
generic function $f(R)$. Starobinsky \cite{15} proposed the first
inflationary model in $f(R)$ which is compatible with anisotropies
of CMBR (Cosmic Microwave Background Radiation). Bamba {\it et al.}
\cite{16} introduced $f(R)$ model which explains inflation and late
cosmic expansion at the same time. There are also other models in
$f(R)$ gravity \cite{17,18} that deal with cosmological constraints
and can resolve some cosmic issues.

Cembranos {\it et al.} \cite{19} investigated spherical dust
collapse in this gravity and showed that collapsing process slows
down due to the contribution of $f(R)$ terms. Sharif and Yousaf
\cite{20} discussed the collapse with metric and Palatini $f(R)$
gravity considering early and late time models. Sharif and Zunaira
\cite{24,25} explored the effects of this gravity on axially
symmetric fluid under shear-free and expansion-free conditions with
the help of structure scalars. Capozziello {\it et al.} \cite{26}
investigated axially symmetric solutions in this gravity using
Noether symmetry approach and discussed physical properties of the
fluid. Rippl {\it et al.} \cite{27} studied the evolution equations
for kinematical variables, electric and magnetic parts of the Weyl
tensor. N\"{a}f and Jetzer \cite{28} described gravitational
radiation of an isolated system in $f(R)$ quadratic model. They used
correspondence between $f(R)$ as well as scalar-tensor theories and
found results inconsistent with GR. The phenomenon of gravitational
waves has also been studied in this gravity \cite{29}.

In order to analyze the influence of higher order curvature terms
($f(R)$ gravity) upon the phenomenon of gravitational radiation, we
explore the evolution of dissipative dust configuration in $f(R)$
gravity. The paper has following format. In the coming section, a
basic formalism of $f(R)$ gravity and the necessary ingredients to
discuss evolution of the fluid are given. Sections \textbf{3} and
\textbf{4} deal with evolution of non-spinning and spinning fluids,
respectively. The last section provides summary of the results
obtained.

\section{Basic Formalism}

The action of $f(R)$ gravity is defined as
\begin{equation}\label{1}
S=\frac{1}{16\pi}\int\sqrt{-g}f(R)d^{4}x+S_{M},
\end{equation}
where $S_{M}=\int\sqrt{-g}L_{M}d^{4}x$ shows the matter action in
which $L_{M}$ represents the matter Lagrangian. The field equations
obtained by varying Eq.(\ref{1}) with respect to the metric tensor
are given by
\begin{equation}\label{2}
R_{\beta\gamma}-\frac{1}{2}Rg_{\beta\gamma}=8\pi
\left\{T_{\beta\gamma}^{(m)}
+\frac{1}{8\pi}T_{\beta\gamma}^{(D)}\right\} = 8\pi
T_{\beta\gamma}^{tot}.
\end{equation}
Here $T_{\beta\gamma}^{(m)}$ represents the energy-momentum tensor
of matter and $T_{\beta\gamma}^{(D)}$ shows the contribution of dark
source terms defined as
\begin{equation}\label{3}
T_{\beta\gamma}^{(D)}=\frac{1}{8\pi}\left\{(1-F)R_{\beta\gamma}
+\frac{g_{\beta\gamma}}{2}(f-R)+
\nabla_{\beta}\nabla_{\gamma}F-g_{\beta\gamma}\Box{F}\right\},
\end{equation}
where $F=\frac{df}{dR}$ and $\Box=\nabla^{\nu}\nabla_{\nu}$. The
density $(\rho^{(D)})$, pressure $(p^{(D)})$ and energy flux
$(q^{(D)}_{\alpha})$ associated with the dark source are given by
\cite{41}
\begin{equation}\label{3}
\rho^{(D)}=T_{\beta\gamma}^{(D)}V^{\beta}V^{\gamma}, \quad
p^{(D)}=\frac{1}{3}T_{\beta\gamma}^{(D)}h^{\beta\gamma},\quad
q^{(D)}_{\alpha}=-T_{\beta\gamma}^{(D)}V^{\beta}h^{\gamma}_{\alpha},
\end{equation}
where $V^{\beta}$ is the four velocity and $h^{\beta\gamma}$ denotes
the projection tensor. The axially symmetric spacetime with
reflection symmetry and geodesic condition is given by \cite{13}
\begin{equation}\label{5}
ds^2=-dt^{2}+B^2(t,r,\theta)\left(dr^{2}+r^{2}d\theta^2\right)
+2\tilde{G}(r,\theta)dtd\theta+C^2(t,r,\theta)d\phi^2.
\end{equation}
The energy-momentum tensor of dust suffering heat dissipation in the
form of radiation is given by
\begin{equation}\label{6}
T_{\beta\gamma}^{(m)}={\rho}V_{\beta}V_{\gamma}+q_{\beta}V_{\gamma}+q_{\gamma}V_{\beta},
\end{equation}
where $\rho$ is the energy density, $q_{\gamma}$ indicates the heat
flux vector and $V_{\gamma}$ shows the four velocity. The four
velocity and unit spacelike vectors in comoving coordinates are
defined as
\begin{eqnarray}\label{7}
V_{\gamma}&=&(-1,0,\tilde{G},0), \quad
K_{\gamma}=(0,B,0,0),\\\nonumber L_{\gamma}&=&(0,0,\sqrt{r^{2}B^{2}
+\tilde{G}^{2}},0), \quad S_{\gamma}=(0,0,0,C).
\end{eqnarray}
These vectors describe a set of orthonormal tetrad and satisfy the
following relations
\begin{eqnarray*}
V_{\gamma}V^{\gamma}&=&-K^{\gamma}K_{\gamma}
=-L^{\gamma}L_{\gamma}=-S^{\gamma}S_{\gamma}=-1, \\
V_{\gamma}K^{\gamma}&=&V^{\gamma}L_{\gamma}=V^{\gamma}S_{\gamma}
=K^{\gamma}L_{\gamma}=K^{\gamma}S_{\gamma}=S^{\gamma}L_{\gamma}=0.
\end{eqnarray*}
Since, $G_{03}$ is zero, so from the field equations, we have
$T_{03}^{tot}=0$ which implies that $q_{3}^{tot}=0$. Thus we can
write
\begin{equation}\label{8}
q_{\alpha}^{tot}= q_{I}^{tot}K_{\alpha}+q_{II}^{tot}L_{\alpha}.
\end{equation}
Also, we have
\begin{equation}\label{9}
q_{\alpha}^{tot}=-T_{\beta\gamma}V^{\beta}V^{\gamma}V_{\alpha}-T_{\alpha\beta}V^{\beta}.
\end{equation}
These two equations yield
\begin{eqnarray}\label{10}
q_{I}^{tot}&=&-\frac{1}{B}T_{01}^{tot}V^{0}=q_{I}-\frac{1}{8\pi
B}T_{01}^{(D)},\\\label{11}
q_{II}^{tot}&=&-\frac{1}{\sqrt{r^{2}B^{2}+\tilde{G}^{2}}}\left\{\tilde{G}T_{00}^{(D)}+
{\sqrt{r^{2}B^{2}+\tilde{G}^{2}}}q_{II}+T_{02}^{(D)}\right\}.
\end{eqnarray}

Now we define kinematical quantities as well as electric and
magnetic parts of the Weyl tensor. The kinematical variables are
very important tools to discuss the evolution of any astrophysical
system. These variables include four acceleration, expansion scalar,
shear tensor, and vorticity tensor. Since we are considering
geodesic fluid, so the four acceleration is zero. The expansion
scalar is given by
\begin{equation}\label{12}
\Theta=V_{;\gamma}^{\gamma}=\left(\frac{2r^{2}B^{2}+\tilde{G}^{2}}
{r^{2}B^{2}+\tilde{G}^{2}}\right)\frac{\dot{B}}{B}+\frac{\dot{C}}{C},
\end{equation}
where dot represents derivative with respect to $t$. The shear
tensor is defined as
\begin{equation}\nonumber
\sigma_{\beta\gamma}=V_{(\beta;\gamma)}+a_{(\beta}V_{\gamma)}-\frac{1}{3}\Theta
h_{\beta\gamma},
\end{equation}
whose non-zero components are $\sigma_{11},~\sigma_{22}$ and
$\sigma_{33}$. In terms of unit spacelike vectors, we can have
\begin{equation}\label{13}
\sigma_{\beta\gamma}=\frac{1}{3}(2\sigma_{I}+\sigma_{II})(K_{\beta}K_{\gamma}
-\frac{1}{3}h_{\beta\gamma})+
\frac{1}{3}(2\sigma_{II}+\sigma_{I})(L_{\beta}L_{\gamma}-\frac{1}{3}h_{\beta\gamma}),
\end{equation}
where
\begin{equation}\label{14}
\sigma_{I}=\left(\frac{r^{2}B^{2}+2\tilde{G}^{2}}{r^{2}B^{2}+
\tilde{G}^{2}}\right)\frac{\dot{B}}{B}-\frac{\dot{C}}{C} ,\quad
\sigma_{II}=\left(\frac{r^{2}B^{2}-\tilde{G}^{2}}{r^{2}B^{2}
+\tilde{G}^{2}}\right)\frac{\dot{B}}{B}-\frac{\dot{C}}{C},
\end{equation}
giving
\begin{equation}\label{15}
\sigma_{I}-\sigma_{II}=\frac{3\tilde{G}^{2}}{r^{2}B^{2}+\tilde{G}^{2}}\frac{\dot{B}}{B}.
\end{equation}
The vorticity tensor is
\begin{equation}\nonumber
\Omega_{\beta\gamma}=V_{[\beta;\gamma]}+a_{[\beta}V_{\gamma]},
\end{equation}
whose non-zero component is $\Omega_{12}=\frac{-\tilde{G}'}{2}$,
implying that
\begin{equation}\label{16}
\Omega_{\beta\gamma}=\Omega(L_{\beta}K_{\gamma}-L_{\gamma}K_{\beta}),
\end{equation}
where $\Omega$ is the vorticity scalar given by
\begin{equation}\label{55}
\Omega=\Omega^{\beta\gamma}\Omega_{\beta\gamma}=\frac{\tilde{G'}}
{2B\sqrt{r^{2}B^{2}+\tilde{G}^{2}}}.
\end{equation}
Here prime denotes differentiation with respect to the radial
coordinate $r$.

The electric and magnetic parts of the Weyl tensor are defined as
\begin{equation}\nonumber
E_{\beta\gamma}=C_{\beta\delta\gamma\lambda}V^{\delta}V^{\lambda},
\quad H_{\beta\gamma}=\frac{1}{2}\eta_{\beta\delta\epsilon\lambda}
C^{\quad \epsilon\lambda}_{\gamma\mu}V^{\delta}V^{\mu},
\end{equation}
where the electric part has three non-zero components
$E_{11},~E_{22}$ and $E_{12}$ while the magnetic part has $H_{13}$
and $H_{23}$. Thus we can write them in terms of unit spacelike
vectors as
\begin{eqnarray}\nonumber
E_{\beta\gamma}&=&\frac{1}{3}(\varepsilon_{II}+2\varepsilon_{I})
\left(K_{\beta}K_{\gamma}-\frac{1}{3}h_{\beta\gamma}\right)+
\frac{1}{3}(\varepsilon_{I}+2\varepsilon_{II})
\left(L_{\beta}L_{\gamma}-\frac{1}{3}h_{\beta\gamma}\right)\\\label{17}&+&
\varepsilon_{KL}\left(K_{\beta}L_{\gamma}+K_{\gamma}L_{\beta}\right),
\\\label{18}
H_{\beta\gamma}&=&H_{1}\left(S_{\beta}K_{\gamma}+S_{\gamma}K_{\beta}\right)+
H_{2}\left(S_{\beta}L_{\gamma}+S_{\gamma}L_{\beta}\right).
\end{eqnarray}
The values of scalars
$\varepsilon_{I},~\varepsilon_{II},~\varepsilon_{KL},~H_{1}$ and
$H_{2}$ in terms of metric functions are given in \cite{24}. The
elementary flatness condition states that a spacetime is locally
isometric to Minkowski spacetime. The regularity condition necessary
for the elementary flatness at the center $(r=0)$ implies that
$\tilde{G}=0$ or it is a regular function of $r$ in the neighborhood
of the center such that it approaches to zero as $r\approx0$
\cite{14,32}. When $\tilde{G}=0$, Eq.(\ref{55}) gives
$\tilde{G}=0\Leftrightarrow\Omega=0$ and in the neighborhood of
center, we may write
\begin{equation}\label{19}
\tilde{G}=\sum_{m=3}^{\infty} \tilde{G}^{(m)}(\theta)~r^{m}.
\end{equation}
Substituting this value in Eq.(\ref{55}), it follows that
\begin{equation}\label{20}
\Omega=\sum_{m=1}^{\infty} \Omega^{(m)}(t,\theta)~r^{m}.
\end{equation}
The condition of elementary flatness \cite{31} implies that
\begin{equation}\label{21}
C\thickapprox r \psi(t,\theta),\quad\text{where}
\quad\psi(t,\theta)\approx B(t,0,\theta).
\end{equation}
Also, in the neighborhood of the center we can write
\begin{equation}\label{aa}
B(t,r,\theta)=\sum_{m=0}^{\infty} B^{(m)}(t,\theta)~r^{m},\quad
C(t,r,\theta)=\sum_{m=1}^{\infty} C^{(m)}(t,\theta)~r^{m}.
\end{equation}
We discuss evolving axially symmetric dissipative dust fluid at the
center and in its vicinity. In the first case, at the center we have
$\Omega=0\Leftrightarrow\tilde{G}=0$, i.e., the fluid is
non-spinning. In the second case, we consider $\tilde{G}$ and
$\Omega$ as regular functions of $r$, i.e., spinning fluid in the
neighborhood of the center as defined in Eqs.(\ref{19}) and
(\ref{20}).

Bel \cite{37} introduced spatial tensors through orthogonal
splitting of the Riemann tensor given as
\begin{equation}\nonumber
Y_{\beta\gamma}=R_{\beta\alpha\gamma\delta}V^{\alpha}V^{\delta},\quad
X_{\beta\gamma}=\frac{1}{2}\eta_{\beta\alpha}^{\quad\mu\nu}R^{*}_{\mu\nu\gamma\delta}
V^{\alpha}V^{\delta},\quad
Z_{\beta\gamma}=\frac{1}{2}\epsilon_{\beta\mu\nu}R_{\delta\gamma}^{\quad\mu\nu}V^{\delta}.
\end{equation}
The scalars obtained from these tensors are called structure scalars
which explain different physical processes during evolution of the
system. we can define super energy and super-Poynting vector using
the above spatial tensors as
\begin{eqnarray}\nonumber
W&=&\frac{1}{2}[X^{\beta\gamma}X_{\beta\gamma}+Y^{\beta\gamma}Y_{\beta\gamma}]+
Z^{\beta\gamma}Z_{\beta\gamma},\\\nonumber
P_{\gamma}&=&\epsilon_{\gamma\mu\nu}(Y_{\delta\rho}Z^{\mu\rho}-
X_{\delta\rho}Z^{\rho\mu})g^{\nu\delta},
\end{eqnarray}
where super energy having dimensions $L^{-4}$ can be interpreted as
the energy per unit area and super-Poynting vector tells about the
state of intrinsic gravitational radiation \cite{38}. Sharif and
Zunaira \cite{24} studied structure scalars in the context of $f(R)$
gravity and evaluated super energy and super-Poynting vector for the
model $R+\xi R^{2}$. We use this super-Poynting vector to confirm
the presence of gravitational radiation in both cases. The
super-Poynting vector in terms of spacelike vectors is given by
\begin{equation}\label{22}
P_{\mu}=P_{I}K_{\mu}+P_{II}L_{\mu}.
\end{equation}
Here the scalars $P_{I}$ and $P_{II}$ describe the flux of super
energy which is a combination of gravitational radiation, heat
dissipation and dark energy as
\begin{eqnarray}\label{23}
P_{I}&=&2H_{1}\varepsilon_{KL}+\frac{2H_{2}}{3}(\varepsilon_{I}+2\varepsilon_{II})+
D_{1},\\\label{24}
P_{II}&=&2H_{2}\varepsilon_{KL}-\frac{2H_{1}}{3}(2\varepsilon_{I}+\varepsilon_{II})+
D_{2},
\end{eqnarray}
where $D_{1}$ and $D_{2}$, given in Appendix {\bf B}, describe the
effects of heat dissipation as well as $f(R)$ gravity. The scalars
$H_{1},~H_{2},~\varepsilon_{I},~\varepsilon_{II}$ and
$\varepsilon_{KL}$, associated with magnetic and electric parts of
the Weyl tensor are linked with gravitational radiation. If the
values of these scalars obtained from evolution equations are
non-zero then the above mentioned terms in the expression of
super-Poyinting vector do not vanish indicating the existence of
gravitational radiation.

For dissipative dust, the evolution equations for the expansion
scalar, shear tensor and vorticity tensor obtained from the Ricci
identities for $f(R)$ gravity are respectively, given by
\begin{eqnarray}\label{1}
\Theta_{;\beta}V^{\beta}+\frac{1}{3}\Theta^{2}
+2\left(\sigma^{2}-\Omega^{2}\right)+\frac{1}{F}\left[ 4\pi\mu
+F_{R}
h^{\lambda\delta}\nabla_{\lambda}\nabla_{\delta}R\right]=0,\\
\nonumber
h^{\alpha}_{\mu}h^{\beta}_{\nu}\sigma_{\alpha\beta;\gamma}V^{\gamma}+
\sigma^{\alpha}_{\mu}\sigma_{\nu\alpha}+
\frac{2}{3}\Theta\sigma_{\mu\nu}-\frac{1}{3}\left(2\sigma^{2}+\Omega^{2}\right)h_{\mu\nu}
+\omega_{\mu}\omega_{\nu}\\\label{2}
+E_{\mu\nu}+\frac{1}{2F}F_{R}\nabla_{\alpha}\nabla_{\beta}R\left
(h^{\alpha}_{\mu}h^{\beta}_{\nu}-
\frac{1}{3}h_{\mu\nu}h^{\alpha\beta}\right)=0, \\\label{3}
h^{\alpha}_{\mu}h^{\beta}_{\nu}\Omega_{\alpha\beta;\gamma}V^{\gamma}+
\frac{2}{3}\Theta\Omega_{\mu\nu}-2\sigma_{\alpha[\mu}\Omega_{\nu]}^{\alpha}=0.
\end{eqnarray}
The constraint equations are
\begin{eqnarray}\label{4}
h_{\mu}^{\nu}\left(\frac{2}{3}\Theta_{;\nu}-
h^{\delta\gamma}\sigma_{\nu\delta;\gamma}\right)+
\eta_{\mu}^{\nu\gamma\delta}V_{\delta}\omega_{\gamma;\nu}-\frac{1}{F}\left[8\pi
q_{\mu}+F_{R}
h_{\mu}^{\nu}(\nabla_{\nu}\nabla_{\gamma})V^{\gamma}\right]=0,\\\label{5}
h_{(\mu}^{\alpha}h_{\nu)\beta}
\left(\sigma_{\alpha\delta}+\Omega_{\alpha\delta}\right)_{;\gamma}
\eta^{\beta\kappa\gamma\delta}V_{\kappa}=H_{\mu\nu}.
\end{eqnarray}
The conservation law gives
\begin{eqnarray}\nonumber\label{6}
&&h_{\mu}^{\nu}q_{\nu;\alpha}V^{\alpha}+\left(\frac{4}{3}\Theta
h_{\mu\nu}+\sigma_{\mu\nu}+\Omega_{\mu\nu}\right)q^{\nu}
+\frac{1}{8\pi}\left[(1-F)R^{\nu}_{\mu}\right.
\\\label{A.6}&+&\left.\nabla_{\nu}\nabla_{\mu}F
+\delta^{\nu}_{\mu}\left(\frac{1}{2}(f-R)-\Box
F\right)\right]_{;\nu}=0.
\end{eqnarray}

The evolution and constraint equations for electric part of the Weyl
tensor are given by
\begin{eqnarray}\nonumber
&&h^{\alpha}_{(\mu}h^{\beta}_{\nu)}E_{\alpha\beta;\delta}V^{\delta}+\Theta
E_{\mu\nu}+h_{\mu\nu}E_{\alpha\beta}\sigma^{\alpha\beta}-
3E_{\alpha(\mu}\sigma_{\nu)}^{\alpha}+
h^{\alpha}_{(\mu}\eta_{\nu)}^{\delta\gamma\kappa}V_{\delta}H_{\gamma\alpha;\kappa}
\\\nonumber&-&
E_{\delta(\mu}\Omega_{\nu)}^{\delta}=\frac{1}{F}\left\{-4\pi\rho
\sigma_{\mu\nu}+\frac{4\pi}{3}q^{\alpha}_{;\alpha}h_{\mu\nu}-4\pi
h^{\alpha}_{(\mu}h^{\beta}_{\nu)}q_{\beta;\alpha}\right\}\\\nonumber
&+& \frac{F_{R}}{2F}h^{\alpha}_{(\mu}h^{\beta}_{\nu)}\left
\{\dot{R}\left(R_{\alpha\beta}-\frac{1}{3}R
g_{\alpha\beta}\right)-(\nabla_{\alpha}R)
R_{\beta\gamma}V^{\gamma}\right.\\\label{7}&+&
\left.\nabla_{\alpha}(\nabla_{\gamma}\nabla_{\beta}R)V^{\gamma}-
(\nabla_{\alpha}\nabla_{\beta}R)^{.}\right\}, \\\nonumber
&&h^{\alpha}_{\mu}h^{\beta\nu}E_{\alpha\beta;\nu}-
\eta_{\mu}^{\delta\beta\kappa}V_{\delta}\sigma^{\gamma}_{\beta}
H_{\kappa\gamma}+3H_{\mu\nu}\omega^{\nu}=\frac{1}{F}
\left\{\frac{8\pi}{3}h^{\nu}_{\mu}\rho_{;\nu}-4\pi\right.
\\\nonumber&\times&\left.\left(\frac{2}{3}\Theta
h^{\nu}_{\mu}-\sigma^{\nu}_{\mu}+3\Omega^{\nu}_{\mu}\right)q_{\nu}\right\}+
\frac{F_{R}}{2F}h_{\mu}^{\alpha}\left\{\dot{R}R_{\alpha}^{\nu}V_{\nu}-
\left(R_{\nu\gamma}V^{\nu}V^{\gamma}+
\frac{1}{3}R\right)\right.\\\label{8}&\times&\left.R_{;\alpha}-(\Box
R)_{;\alpha}-(\nabla_{\alpha}\nabla_{\nu}R)^{.}V^{\nu}+
h^{\nu}_{\gamma}\nabla_{\alpha}(\nabla_{\nu}\nabla^{\gamma}R)\right\},
\end{eqnarray}
where $F=\frac{df}{dR}$ and $F_{R}=\frac{d^{2}f}{dR^{2}}$. The value
of Ricci scalar $R$, for the spacetime given in Eq.(\ref{5}) is
calculated as
\begin{eqnarray}\nonumber
R&=&\frac{1}{r^{2}B^{2}+\tilde{G}^{2}}[\frac{2}{C}(C_{,\theta\theta}-
r^{2}B^{2}\ddot{C}+2\tilde{G}\dot{C}_{,\theta}
-2r^{2}B\dot{C}\dot{B}+r
C')-\frac{2}{B^{2}C}\\\nonumber&\times&[C''(r^{2}B^{2}+\tilde{G}^{2})-\tilde{G}\tilde{G}'C'
-\tilde{G}C
\tilde{G}'']+\frac{2}{B}(2\tilde{G}\dot{B}_{,\theta}+r^{2}B''+B_{,\theta\theta})\\\nonumber&-&
\frac{2\tilde{G}^{2}B'C'}{B^{3}C}]-\frac{1}{2B^{3}C(r^{2}B^{2}+\tilde{G}^{2})^{2}}
[4\tilde{G}C\tilde{G}'B'(3r^{2}B^{2}+\tilde{G}^{2})+8r^{2}B^{4}\dot{C}\dot{B}
\\\nonumber&\times&(r^{2}B^{2}+2\tilde{G}^{2})+(r^{2}B^{2}+\tilde{G}^{2})(4B^{3}r^{2}C\dot{B}^{2}
-4B^{2}C C'B')+4B^{3}C
r^{2}\\\nonumber&\times&(r^{2}B'^{2}+B_{,\theta}^{2})+4\tilde{G}_{,\theta}(B^{3}\tilde{G}C_{,\theta}
-B^{5}r^{2}\dot{C}-B^{4}r^{2}C\dot{B}+B^{2}C\tilde{G}B_{,\theta})\\\label{ss}&+&
4B_{,\theta}B^{2}\tilde{G}(2Cr^{2}B\dot{B}-\tilde{G}C_{,\theta}-\tilde{G}^{2}\dot{C})+
8\tilde{G}CrB^{3}G'-4B^{3}C\tilde{G}^{2}].
\end{eqnarray}
From Eqs.(\ref{19}) and (\ref{aa}), the behavior of $R$ near the
center is given by
\begin{equation}\label{cc}
R(t,r,\theta)=\sum_{m=-2}^{\infty} R^{(m)}(t,\theta)~r^{m}.
\end{equation}
It gives $R\rightarrow\infty$ as $r\rightarrow0$. This implies that
there is a singularity at $r=0$ or at the center of considered
configuration (which makes the curvature scalar infinite) \cite{6}.
In this paper, we use the $f(R)$ model given as
\begin{equation}\nonumber
f(R)=\lambda R+\xi R^{2},
\end{equation}
where $\lambda$ and $\xi$ are positive real numbers. In the limit
$\lambda\rightarrow1$ and $\xi\rightarrow0$, this reduces to GR. Any
model in which $f(R)\propto R^{2}$ can explain the inflationary
scenario but cannot tell about the recent cosmic expansion. However,
the inclusion of linear term in $R$ causes the inflation to end when
$ R^{2}$ term is smaller than the linear term (this occurs by a
reheating stage in which gravitational particles are produced due to
oscillations in $R$) \cite{39}. Consequently, it can lead to
expansion of the universe. This model being quadratic can also
explain dark matter and dark energy \cite{40}. This reduces to
Starobinsky inflationary model when $\lambda\rightarrow1$, which
estimates nearly flat spectrum of gravitational waves and is also
consistent with the temperature anisotropies measured by CMBR
\cite{39}.

\section{Non-Spinning Dissipative Dust}

In this section, we deal with the non-spinning case. Using the
condition $\tilde{G}=0\Leftrightarrow\Omega=0$, Eqs.(\ref{12}) and
(\ref{14}) reduce to the following form
\begin{equation}\label{25}
\Theta=2\frac{\dot{B}}{B}+\frac{\dot{C}}{C}~,\quad
\sigma_{I}=\sigma_{II}=\tilde{\sigma}=\frac{\dot{B}}{B}+\frac{\dot{C}}{C}~.
\end{equation}
The general evolution equations obtained through Ricci and Bianchi
identities are given in Appendix \textbf{A} while the contracted
equations are in Appendix \textbf{B}. The equations
(\ref{B.5})-(\ref{B.8}) along with $\Omega=0=\tilde{G}$ give
\begin{eqnarray}\label{26}
\frac{1}{3B}\left\{2\Theta^{'}-\tilde{\sigma}'-\tilde{\sigma}\frac{3C^{'}}{C}\right\}=
\frac{1}{(\lambda+2\xi R)}\left\{8\pi q_{I}+2 \xi
K^{\nu}(\nabla_{\nu}\nabla_{\lambda}R)V^{\lambda}\right\},
\\\label{27}
\frac{1}{3rB}\left\{2\Theta_{,\theta}-\tilde{\sigma}_{,\theta}-\tilde{\sigma}\frac{3C_{,\theta}}{C}\right\}=
\frac{1}{(\lambda+2\xi R)}\left\{8\pi q_{II}+2 \xi
L^{\nu}(\nabla_{\nu}\nabla_{\lambda}R)V^{\lambda}\right\},\\\label{28}
H_{1}=-\frac{\tilde{\sigma}}{2rB}\left(\frac{\tilde{\sigma}_{,\theta}}{\tilde{\sigma}}
+\frac{C_{,\theta}}{C}\right)=-\frac{(\tilde{\sigma}C)_{,\theta}}{2rBC},
\\\label{29}
H_{2}=\frac{\tilde{\sigma}}{2B}\left(\frac{\tilde{\sigma}^{'}}{\tilde{\sigma}}
+\frac{C^{'}}{C}\right)=\frac{(\tilde{\sigma}C)'}{2BC}.
\end{eqnarray}
Using Eq.(\ref{25}) in Eqs.(\ref{26}) and (\ref{27}), we obtain
\begin{equation}\label{30}
q_{I}=-\frac{1}{8\pi}\left\{2\xi
K^{\nu}(\nabla_{\nu}\nabla_{\kappa})V^{\kappa}\right\}+\frac{(\lambda+2\xi
R)}{8\pi
B}\left\{\left(\frac{\dot{B}}{B}\right)'-\frac{\dot{B}}{B}\frac{C'}{C}+\frac{\dot{C}'}{C}\right\},
\end{equation}
\begin{equation}\label{31}
q_{II}=-\frac{1}{8\pi}\left\{2\xi
L^{\nu}(\nabla_{\nu}\nabla_{\kappa})V^{\kappa}\right\}+\frac{(\lambda+2\xi
R)}{8\pi rB}\left\{\left(\frac{\dot{B}}{B}\right)_{,\theta}
-\frac{\dot{B}}{B}\frac{C_{,\theta}}{C}+\frac{\dot{C}_{,\theta}}{C}\right\}.
\end{equation}
These show the presence of heat flux as well as the effect of $f(R)$
curvature terms on dissipation. Using Eqs.(\ref{30}) and (\ref{31})
in (\ref{28}) and (\ref{29}), we obtain
\begin{eqnarray}\label{32}
H_{1}&=&-\frac{1}{rB}\left(\frac{\dot{B}}{B}\right)_{,\theta}+\frac{1}{\lambda+2\xi
R}\left\{4\pi q_{II}+\xi
L^{\nu}(\nabla_{\nu}\nabla_{\kappa})V^{\kappa}\right\},
\\\label{33}
H_{2}&=&-\frac{1}{B}\left(\frac{\dot{B}}{B}\right)'-\frac{1}{\lambda+2\xi
R}\left\{4\pi q_{I}+\xi
K^{\nu}(\nabla_{\nu}\nabla_{\kappa})V^{\kappa}\right\}.
\end{eqnarray}
These equations indicate that $H_{1}$ and $H_{2}$ have the
contribution of heat flux and $f(R)$ gravity. If we assume $B$ and
$C$ explicit functions of $t$ and $(r,\theta)$ as
$B(t,r,\theta)=T(t)\tilde{B}(r,\theta)$ and
$C(t,r,\theta)=T(t)\tilde{C}(r,\theta)$ then
Eqs.(\ref{30})-(\ref{33}) become
\begin{eqnarray}\label{a1}
q_{I}&=&-\frac{1}{8\pi}\left\{2\xi
K^{\nu}(\nabla_{\nu}\nabla_{\kappa})V^{\kappa}\right\},\\\label{a2}
q_{II}&=&-\frac{1}{8\pi}\left\{2\xi
L^{\nu}(\nabla_{\nu}\nabla_{\kappa})V^{\kappa}\right\},\\\label{a3}
H_{1}&=&\frac{1}{\lambda+2\xi R}\left\{4\pi q_{II}+\xi
L^{\nu}(\nabla_{\nu}\nabla_{\kappa})V^{\kappa}\right\},
\\\label{a4}
H_{2}&=&-\frac{1}{\lambda+2\xi R}\left\{4\pi q_{I}+\xi
K^{\nu}(\nabla_{\nu}\nabla_{\kappa})V^{\kappa}\right\}.
\end{eqnarray}
In the limit $\xi\rightarrow0$, we obtain the results of GR. From
Eq.(\ref{cc}) we see that when $r\rightarrow0$, $R\rightarrow\infty$
and Eq.(\ref{a3}) and (\ref{a4}) give
\begin{equation}\nonumber
H_{1}=0,~H_{2}=0.
\end{equation}
Thus Eqs.(\ref{22})-(\ref{24}) indicate that the scalars of
super-Poynting vector and hence the super-Poynting vector contains
no contribution due to gravitational radiation. It is mentioned here
that the non-spinning dissipative dust is also not gravitationally
radiating in GR. To find the values of
$\varepsilon_{I},~\varepsilon_{II}$ and $\varepsilon_{KL}$, we
consider Eqs.(\ref{B.9}) and (\ref{B.10}) given by
\begin{equation}\label{34}
\frac{1}{3}\dot{\varepsilon}_{I}+\frac{1}{3}\varepsilon_{I}\Theta+
\frac{1}{3}\varepsilon_{II}\tilde{\sigma}=0,
\end{equation}
\begin{equation}\label{35}
\frac{1}{3}\dot{\varepsilon}_{II}+\frac{1}{3}\varepsilon_{II}\Theta+
\frac{1}{3}\varepsilon_{I}\tilde{\sigma}=0,
\end{equation}
where we have used the values of $H_{1}$, $H_{2}$ and $R$ at
$r\rightarrow0$. Solving the above two equations simultaneously, we
obtain
\begin{eqnarray}\label{36}
\varepsilon_{I}&=&\frac{1}{2B}\left\{\frac{1}{B^{2}C^{2}}+1\right\},\\\label{37}
\varepsilon_{II}&=&\frac{1}{2B}\left\{\frac{1}{B^{2}C^{2}}-1\right\}.
\end{eqnarray}
As $C\rightarrow0$ when $r\rightarrow0$ implying that
$\varepsilon_{I}$ and $\varepsilon_{II}$ approaches to infinity.
Similarly, the value of $\varepsilon_{KL}$ obtained from
Eq.(\ref{B.12}) is given by
\begin{eqnarray}\label{38}
\varepsilon_{KL}=\frac{1}{B}.
\end{eqnarray}

\section{Spinning Dissipative Dust}

In this case, we take non-zero vorticity scalar (spinning fluid),
which is a regular function of $r$ such that it vanishes at the
center. Moreover, all geometrical and physical variables are regular
at $r\approx 0$. Using Eqs.(\ref{19}) and (\ref{aa}) in
Eq.(\ref{15}), it follows that
\begin{equation}\label{39}
\sigma_{I}-\sigma_{II}=\sum_{m=4}^{\infty}\left[\sigma_{I}^{(m)}(t,\theta)
-\sigma_{II}^{(m)}(t,\theta)\right]~r^{m}.
\end{equation}
Similarly, from Eq.(\ref{12}) we can write
\begin{equation}\label{39}
\Theta=\sum_{m=0}^{\infty}\Theta^{(m)}(t,\theta) r^{m}.
\end{equation}
Contraction of Eq.(\ref{B.14}) with $K^{\mu}$ and $L^{\mu}$ give the
following equations
\begin{eqnarray}\label{40}
\Omega q_{II}+\frac{\kappa T^{'}}{\tau
B}+\left\{\frac{1}{\tau}+\frac{1}{2}D_{t}\left[\ln
\left(\frac{\tau}{\kappa
T^{2}}\right)\right]-\frac{5}{6}\Theta\right\}q_{I}-\frac{q_{I}\sigma_{I}}{3}
+ D_{3}=0,\\\label{41} -\Omega q_{I}+\frac{\kappa
L^{\mu}T_{,\mu}}{\tau}+\left\{\frac{1}{\tau}+\frac{1}{2}D_{t}\left[\ln
\left(\frac{\tau}{\kappa
T^{2}}\right)\right]-\frac{5}{6}\Theta\right\}q_{II}-\frac{q_{II}\sigma_{II}}{3}
+ D_{4}=0,
\end{eqnarray}
where $D_{t}$ represents covariant derivative with respect to time
and $\tau$ is the relaxation time. Here $D_{3}$ and $D_{4}$ indicate
the terms due to $f(R)$ extra degrees of freedom given in
(\ref{B.16}) and (\ref{B.17}), respectively. From Eqs.(\ref{B.16})
and (\ref{B.17}), the behavior of $D_{3}$ and $D_{4}$ is given by
\begin{equation}\label{39}
D_{3}=\sum_{m=-5}^{\infty}D_{3}^{(m)}(t,\theta)
r^{m},~D_{4}=\sum_{m=-5}^{\infty}D_{4}^{(m)}(t,\theta) r^{m}
\end{equation}
Since we have assumed that all the variables are regular, so is the
dissipation scalars $q_{I}$ and $q_{II}$. Let us take
\begin{eqnarray}\label{42}
q_{I}=\sum_{m=0}^{\infty}q^{(m)}_{I}(t,\theta)~r^{m},\quad
q_{II}=\sum_{m=0}^{\infty}q^{(m)}_{II}(t,\theta)~r^{m},
\end{eqnarray}
near the center. Considering the terms of $r^{(0)}$ in
Eq.(\ref{40}), we have
\begin{eqnarray}\label{43}
\left\{\frac{1}{\tau}+\frac{1}{2}D_{t}\left[\ln
\left(\frac{\tau}{\kappa
(T^{(0)})^{2}}\right)\right]-\frac{5}{6}\Theta^{(0)}\right\}q^{(0)}_{I}(t,\theta)
+ D^{(0)}_{3}(t,\theta)\\\nonumber+\frac{\kappa}{\tau
B^{(0)}}T^{(1)}(t,\theta)=0,
\end{eqnarray}
where it is assumed that in the neighborhood of center, $T$
(temperature) has the same behavior as $q_{I}$ and $q_{II}$, i.e.,
\begin{equation}\label{39}
T=\sum_{m=0}^{\infty}T^{(m)}(t,\theta) r^{m}.
\end{equation}
Similarly, the coefficients of $r^{(0)}$ in Eq.(\ref{41}) give
\begin{eqnarray}\label{44}
\left\{\frac{1}{\tau}+\frac{1}{2}D_{t}\left[\ln
\left(\frac{\tau}{\kappa
(T^{(0)})^{2}}\right)\right]-\frac{5}{6}\Theta^{(0)}\right\}q^{(0)}_{II}(t,\theta)
+ D^{(0)}_{4}(t,\theta)\\\nonumber+\frac{\kappa}{\tau
B^{(0)}}T^{(0)}_{,\theta}(t,\theta)=0.
\end{eqnarray}
From the above equations, it is clear that $q^{(0)}_{I}(t,\theta)$
and $q^{(0)}_{II}(t,\theta)$ are non-zero and contain the effects of
$f(R)$ terms. Consequently, when $r\approx0$, the dissipation
scalars are non-zero showing that there is heat dissipation at the
center.

Next excluding the singularities of the scalars
$\varepsilon_{I},~\varepsilon_{II},~\varepsilon_{KL},~H_{1}$ and
$H_{2}$ at the center, we may consider
\begin{eqnarray}\label{45}
\varepsilon_{I}=\sum_{m=0}^{\infty}\varepsilon^{(m)}_{I}(t,\theta)
~r^{m}, \quad
\varepsilon_{II}=\sum_{m=0}^{\infty}\varepsilon^{(m)}_{II}(t,\theta)
~r^{m},\quad
\varepsilon_{KL}=\sum_{m=0}^{\infty}\varepsilon^{(m)}_{KL}(t,\theta)
~r^{m},\\\label{46}
H_{1}=\sum_{m=0}^{\infty}H^{(m)}_{1}(t,\theta)~r^{m},\quad
H_{2}=\sum_{m=0}^{\infty}H^{(m)}_{2}(t,\theta)~r^{m}.
\end{eqnarray}
Using these expressions in Eqs.(\ref{B.1})-(\ref{B.3}) and
comparison of the coefficients of $r^{(0)}$ on both sides
respectively, give
\begin{eqnarray}\label{47}
\varepsilon^{(0)}_{I}(t, \theta)=-D^{(0)}_{5}(t,\theta),\\\label{48}
\varepsilon^{(0)}_{KL}(t,
\theta)=-D^{(0)}_{6}(t,\theta),\\\label{49}
\varepsilon^{(0)}_{II}(t, \theta)=-D^{(0)}_{7}(t,\theta),
\end{eqnarray}
where $D_{5},~D_{6}$ and $D_{7}$ describe the effects of $f(R)$
curvature terms given in Appendix \textbf{B}. The behavior of these
terms near the center is given by
\begin{eqnarray}\nonumber
D_{5}=\sum_{m=-2}^{\infty}D^{(m)}_{5}(t,\theta) ~r^{m}, \quad
D_{6}=\sum_{m=-2}^{\infty}D^{(m)}_{6}(t,\theta) ~r^{m},\quad
D_{7}=\sum_{m=-2}^{\infty}D^{(m)}_{7}(t,\theta) ~r^{m}.
\end{eqnarray}
From Eqs.(\ref{47})-(\ref{49}), it is obvious that scalars of the
electric part of the Weyl tensor do not vanish at the center because
their zeroth order coefficients are non-zero. Also, $D_{5},~D_{6}$
and $D_{7}$ approach to zero as $\xi\rightarrow0$, implying that
$D_{5}^{(0)},~D_{6}^{(0)}$ and $D_{7}^{(0)}$ become zero which in
turn give $\varepsilon^{(0)}_{I}$, $\varepsilon^{(0)}_{KL}$ and
$\varepsilon^{(0)}_{II}$ zero. Further assuming that these scalars
are of class $C^{\omega}$, we have zero value of $\varepsilon_{I}$,
$\varepsilon_{KL}$ and $\varepsilon_{II}$ for the whole
configuration. Thus Eqs.(24) and (25) show that the matter is not
gravitationally radiating, which is the case of GR. Comparing the
coefficients of $r^{(0)}$, Eqs.(\ref{B.9}) and (\ref{B.10}) imply
that
\begin{eqnarray}\label{50}
-H^{(0)}_{1,\theta}-H^{(0)}_{1}\frac{C^{(1)}_{,\theta}}{C^{(1)}}+
\frac{H^{(0)}_{2}}{B^{(0)}}-H^{(0)}_{2}=D_{8}^{(0)},
\\\label{51}
H^{(0)}_{2}+H^{(0)}_{1}[\frac{C^{(1)}_{,\theta}}{C^{(1)}}-\frac{B^{(0)}_{,\theta}}{B^{(0)}}]=
D_{9}^{(0)}.
\end{eqnarray}
The terms $D_{8}$ and $D_{9}$ are given in Appendix \textbf{B} and
their behavior in the neighborhood of the center can be written as
\begin{eqnarray}\nonumber
D_{8}=\sum_{m=-2}^{\infty}D^{(m)}_{8}(t,\theta) ~r^{m}, \quad
D_{9}=\sum_{m=-2}^{\infty}D^{(m)}_{9}(t,\theta) ~r^{m}.
\end{eqnarray}

Using the value of $H_{2}^{(0)}$ from Eq.(\ref{51}) into (\ref{50}),
we obtain
\begin{eqnarray}\label{52}
H^{(0)}_{1,\theta}+H^{(0)}_{1}\left[\frac{C^{(1)}_{,\theta}}{C^{(1)}}
+(1-\frac{1}{B^{(0)}})\left\{\frac{C^{(1)}_{,\theta}}{C^{(1)}}
-\frac{B^{(0)}_{,\theta}}{B^{(0)}}\right\}\right]
=-D_{8}^{(0)}-D_{9}^{(0)},
\end{eqnarray}
which is a first order linear differential equation and its solution
is given by
\begin{equation}\label{53}
H^{(0)}_{1}=-C^{(1)}\left(\frac{C^{(1)}}{B^{(0)}}\right)^{(1-\frac{1}{B^{(0)}})}\int
C^{(1)}\left(\frac{C^{(1)}}{B^{(0)}}\right)^{(1-\frac{1}{B^{(0)}})}(D_{8}^{(0)}+D_{9}^{(0)})~d\theta.
\end{equation}
Substituting $H^{(0)}_{1}$ in Eq.(\ref{51}), we obtain the value of
$H^{(0)}_{2}$. Thus $H^{(0)}_{1},~H^{(0)}_{2}$
$\varepsilon_{I}^{(0)},~\varepsilon_{II}^{(0)}$ and
$\varepsilon_{KL}^{(0)}$ are non-zero implying that the scalars
$H_{1},~H_{2},~\varepsilon_{I},~\varepsilon_{II}$ and
$\varepsilon_{KL}$  do not vanish when $r\approx0$. Assuming that
the scalars defined in Eqs.(\ref{42}), (\ref{45}) and (\ref{46}) are
of class $C^{\omega}$, thus we have non-vanishing values of these
scalars for the whole configuration. Hence the super-Poynting vector
shows that the spinning dust is producing gravitational radiation in
contrast to GR where the fluid is not gravitationally radiating for
this case.

\section{Final Remarks}

The process of gravitational radiation produced by self-gravitating
objects has attracted many researchers due to their fascinating
features. If we could observe these waves it may open new horizons
to solve different cosmic issues. This paper is devoted to study the
gravitational radiation produced by dissipative dust cloud in the
background of $f(R)$ gravity. It is assumed that the matter
configuration is axially symmetric and geodesic. We have discussed
the values of scalars in super-Poynting vector to assure the
presence of gravitational radiation. For this purpose, we have used
evolution equations of non-spinning and spinning fluids. In the
non-spinning case, we have found that the scalars $q_{I}$ and
$q_{II}$ are non-zero at $r=0$ leading to non-zero heat dissipation.
This implies that there is heat flux at the center in the form of
radiation during the evolution process. Equations (\ref{a3}) and
(\ref{a4}) indicate that the scalars associated with magnetic part
of the Weyl tensor are zero. It turns out that geodesic dust with
dissipation do not produce gravitational radiation at the center.

In the second case (spinning fluid), a combination of conservation
and transport equations imply that the values of dissipation scalars
are non-zero when $r\approx0$. Similarly, near the axis of symmetry,
we have discussed the coefficients of $r^{m}$ in Eqs.(\ref{45}) and
(\ref{46}) and found that the magnetic as well as electric parts of
the Weyl tensor are non-zero. In this case, all scalars
$q_{I},~q_{II},~\varepsilon_{I},~\varepsilon_{II},~\varepsilon_{KL},~H_{1}$
and $H_{2}$ depend upon $f(R)$ model. Thus we conclude that the
matter under consideration is gravitationally radiating for
non-vanishing vorticity in the framework of $f(R)$ gravity. It is
worthwhile to mention here that our results reduce to GR in the
limit $\xi\rightarrow0$ \cite{14}. For non-spinning case,
Eqs.(\ref{a1})-(\ref{a4}) give zero values of $H_{1}$ and $H_{2}$ in
the limit $\xi\rightarrow0$ while for spinning case,
Eqs.(\ref{47})-(\ref{49}) yield zero value of $\varepsilon_{I}$,
$\varepsilon_{KL}$ and $\varepsilon_{II}$ implying that the
gravitational part of super-Poynting vector is equal to zero in both
cases. In GR, both cases imply that the matter under consideration
could not act as a source of gravitational radiation.

Thus for a spinning celestial object having axially symmetry there
are no gravitational waves at the center (because the center is on
spinning axis). Also, there is a singularity at the center. The
gravitational radiation is produced in the spacetime fabric due to
the disturbance followed by the spinning of celestial object.
Finally, we can conclude that an astrophysical object having axially
symmetry consisting of dissipative dust can emit gravitational
radiation in the framework of $f(R)$ gravity. This is significant as
there is no gravitational radiation for axially symmetric spinning
celestial object in GR.

\vspace{0.3cm}

\renewcommand{\theequation}{A\arabic{equation}}
\setcounter{equation}{0}
\section*{Appendix A}

The time propagation equations for the expansion scalar, shear
tensor and vorticity tensor obtained from the Ricci identities in
$f(R)$ gravity model $f(R)=\lambda R+\xi R^{2}$ are respectively,
given by
\begin{eqnarray}\label{A.1}
\Theta_{;\beta}V^{\beta}+\frac{1}{3}\Theta^{2}
+2\left(\sigma^{2}-\Omega^{2}\right)+\frac{1}{(\lambda+2\xi
R)}\left[ 4\pi\mu +2\xi
h^{\lambda\delta}\nabla_{\lambda}\nabla_{\delta}R\right]=0,\\
\nonumber
h^{\alpha}_{\mu}h^{\beta}_{\nu}\sigma_{\alpha\beta;\gamma}V^{\gamma}+
\sigma^{\alpha}_{\mu}\sigma_{\nu\alpha}+
\frac{2}{3}\Theta\sigma_{\mu\nu}-\frac{1}{3}\left(2\sigma^{2}+\Omega^{2}\right)h_{\mu\nu}
+\omega_{\mu}\omega_{\nu}\\\label{A.2}
+E_{\mu\nu}+\frac{1}{(\lambda+2\xi
R)}\xi\nabla_{\alpha}\nabla_{\beta}R\left(h^{\alpha}_{\mu}h^{\beta}_{\nu}-
\frac{1}{3}h_{\mu\nu}h^{\alpha\beta}\right)=0, \\\label{A.3}
h^{\alpha}_{\mu}h^{\beta}_{\nu}\Omega_{\alpha\beta;\gamma}V^{\gamma}+
\frac{2}{3}\Theta\Omega_{\mu\nu}-2\sigma_{\alpha[\mu}\Omega_{\nu]}^{\alpha}=0.
\end{eqnarray}
The constraint equations are
\begin{eqnarray}\label{A.4}
h_{\mu}^{\nu}\left(\frac{2}{3}\Theta_{;\nu}-h^{\delta\gamma}\sigma_{\nu\delta;\gamma}\right)+
\eta_{\mu}^{\nu\gamma\delta}V_{\delta}\omega_{\gamma;\nu}-\frac{1}{(\lambda+2\xi
R)}\left[8\pi q_{\mu}+2\xi
h_{\mu}^{\nu}(\nabla_{\nu}\nabla_{\gamma})V^{\gamma}\right]=0.\\\label{A.5}
h_{(\mu}^{\alpha}h_{\nu)\beta}
\left(\sigma_{\alpha\delta}+\Omega_{\alpha\delta}\right)_{;\gamma}
\eta^{\beta\kappa\gamma\delta}V_{\kappa}=H_{\mu\nu}.
\end{eqnarray}
The conservation law gives
\begin{eqnarray}\nonumber\label{A.6}
&&h_{\mu}^{\nu}q_{\nu;\alpha}V^{\alpha}+\left(\frac{4}{3}\Theta
h_{\mu\nu}+\sigma_{\mu\nu}+\Omega_{\mu\nu}\right)q^{\nu}
+\frac{1}{8\pi}\left[\left\{\nabla^{\nu}\nabla_{\mu}(\lambda+2\xi R)
-2\xi R
R^{\nu}_{\mu}\right\}_{;\nu}\right.\\\label{A.6}&-&\left.\delta^{\nu}_{\mu}\left\{2\xi(\Box
R)_{;\nu}+\frac{1}{2}(\xi R^{2})_{;\nu}\right\}\right]=0.
\end{eqnarray}
The evolution and constraint equations for electric part of the Weyl
tensor are given by
\begin{eqnarray}\nonumber
&&h^{\alpha}_{(\mu}h^{\beta}_{\nu)}E_{\alpha\beta;\delta}V^{\delta}+\Theta
E_{\mu\nu}+h_{\mu\nu}E_{\alpha\beta}\sigma^{\alpha\beta}-3E_{\alpha(\mu}\sigma_{\nu)}^{\alpha}+
h^{\alpha}_{(\mu}\eta_{\nu)}^{\delta\gamma\kappa}V_{\delta}H_{\gamma\alpha;\kappa}\\\nonumber&-&
E_{\delta(\mu}\Omega_{\nu)}^{\delta}=\frac{1}{(\lambda+2\xi
R)}\left\{-4\pi\rho
\sigma_{\mu\nu}+\frac{4\pi}{3}q^{\alpha}_{;\alpha}h_{\mu\nu}-4\pi
h^{\alpha}_{(\mu}h^{\beta}_{\nu)}q_{\beta;\alpha}\right\}\\\nonumber
&+& \frac{\xi}{(\lambda+2\xi
R)}h^{\alpha}_{(\mu}h^{\beta}_{\nu)}\left\{\dot{R}\left(R_{\alpha\beta}-\frac{1}{3}R
g_{\alpha\beta}\right)-(\nabla_{\alpha}R)
R_{\beta\gamma}V^{\gamma}\right.\\\label{A.7}&+&\left.\nabla_{\alpha}(\nabla_{\gamma}\nabla_{\beta}R)V^{\gamma}-
(\nabla_{\alpha}\nabla_{\beta}R)^{.}\right\}, \\\nonumber
&&h^{\alpha}_{\mu}h^{\beta\nu}E_{\alpha\beta;\nu}-
\eta_{\mu}^{\delta\beta\kappa}V_{\delta}\sigma^{\gamma}_{\beta}
H_{\kappa\gamma}+3H_{\mu\nu}\omega^{\nu}=\frac{1}{(\lambda+2\xi
R)}\left\{\frac{8\pi}{3}h^{\nu}_{\mu}\rho_{;\nu}-4\pi\right.
\\\nonumber&\times&\left.\left(\frac{2}{3}\Theta
h^{\nu}_{\mu}-\sigma^{\nu}_{\mu}+3\Omega^{\nu}_{\mu}\right)q_{\nu}\right\}+\frac{\xi}{(\lambda+2\xi
R)}h_{\mu}^{\alpha}\left\{\dot{R}R_{\alpha}^{\nu}V_{\nu}-\left(R_{\nu\gamma}V^{\nu}V^{\gamma}+
\frac{1}{3}R\right)\right.\\\label{A.8}&\times&\left.R_{;\alpha}-(\Box
R)_{;\alpha}-(\nabla_{\alpha}\nabla_{\nu}R)^{.}V^{\nu}+
h^{\nu}_{\gamma}\nabla_{\alpha}(\nabla_{\nu}\nabla^{\gamma}R)\right\}.
\end{eqnarray}

\vspace{0.3cm}

\renewcommand{\theequation}{B\arabic{equation}}
\setcounter{equation}{0}
\section*{Appendix B}

Here we contract the equations in appendix A with spacelike vectors
to get scalar equations. Equation (\ref{A.1}) remains the same.
Contraction of Eq.(\ref{A.2}) with $K^{\mu}K^{\nu},~K^{\mu}L^{\nu}$
and with $L^{\mu}L^{\nu}$ gives
\begin{eqnarray}\nonumber
&&\sigma_{I,\lambda}V^{\lambda}+\frac{1}{9}\sigma_{I}^{2}+\frac{2}{3}\Theta
\sigma_{I}-\frac{2}{9}\sigma_{II}\left(\sigma_{I}+\sigma_{II}\right)+
\varepsilon_{I}+\frac{3\xi}{(\lambda+2\xi R)}\\\label{B.1}
&\times&\left(K^{\alpha}K^{\beta}-\frac{1}{3}h_{\mu\nu}K^{\mu}K^{\nu}h^{\alpha\beta}\right)
\nabla_{\alpha}\nabla_{\beta}R=\Omega^{2}, \\\nonumber
&&\frac{1}{3}\left(\sigma_{I}-\sigma_{II}\right)\Omega+\varepsilon_{KL}+\frac{\xi}{(\lambda+2\xi
R)}\left(K^{\alpha}L^{\beta}\right.\\\label{B.2}
&-&\left.\frac{1}{3}h_{\mu\nu}K^{\mu}L^{\nu}h^{\alpha\beta}\right)
\nabla_{\alpha}\nabla_{\beta}R=0, \\\nonumber
&&\sigma_{II,\lambda}V^{\lambda}+\frac{1}{9}\sigma_{II}^{2}+\frac{2}{3}\Theta
\sigma_{II}-\frac{2}{9}\sigma_{I}\left(\sigma_{I}+\sigma_{II}\right)
+\varepsilon_{II}+\frac{3\xi}{(\lambda+2\xi
R)}\\\label{B.3}&\times&\left(L^{\alpha}L^{\beta}-
\frac{1}{3}h_{\mu\nu}L^{\mu}L^{\nu}h^{\alpha\beta}\right)
\nabla_{\alpha}\nabla_{\beta}R=\Omega^{2}.
\end{eqnarray}
Contracting Eq.(\ref{A.3}) with $K^{\mu}L^{\nu}$ yields
\begin{eqnarray}\label{B.4}
\Omega_{,\lambda}V^{\lambda}+\frac{1}{3}\left(2\Theta+\sigma_{I}+\sigma_{II}\right)\Omega=0.
\end{eqnarray}
Contraction of Eq.(\ref{A.4}) with $K^{\mu}$ and $L^{\mu}$ gives
\begin{eqnarray}\nonumber
&&\frac{-1}{\sqrt{r^{2}B^{2}+\tilde{G}^{2}}}\left\{\Omega_{,\theta}+\tilde{G}\dot{\Omega}+
\Omega\left(\frac{\tilde{G}\dot{C}}{C}+\frac{C_{,\theta}}{C}\right)\right\}+
\frac{1}{3B}\left[2\Theta^{'}-\sigma_{I}'
\right.\\\nonumber&-&\left.\sigma_{I}\left\{\frac{2C^{'}}{C}+
\frac{\left(r^{2}B^{2}+\tilde{G}^{2}\right)^{'}}{2\left(r^{2}B^{2}+\tilde{G}^{2}\right)}\right\}
-\sigma_{II}\left\{\frac{2C^{'}}{C}-
\frac{\left(r^{2}B^{2}+\tilde{G}^{2}\right)^{'}}{2\left(r^{2}B^{2}+
\tilde{G}^{2}\right)}\right\}\right]
\\\label{B.5}&=&\frac{1}{\lambda+2\xi R}\left\{8\pi q_{I}+2\xi
K^{\nu}\nabla_{\nu}\nabla_{\lambda}V^{\lambda}\right\}, \\\nonumber
&&\frac{1}{B}\left(\Omega^{'}+\Omega\frac{C^{'}}{C}\right)+
\frac{1}{3(\sqrt{r^{2}B^{2}+\tilde{G}^{2}})}
\left[\left(2\Theta-\sigma_{II}\right)_{,\theta}+\tilde{G}
\left(2\Theta-\sigma_{II}\right)\right.\\\nonumber&+&\left.
\sigma_{I}\left\{\frac{B_{,\theta}}{B}-\frac{C_{,\theta}}{C}+
\tilde{G}\left(\frac{\dot{B}}{B}-\frac{\dot{C}}{C}\right)\right\}-
\sigma_{II}\left\{\frac{B_{,\theta}}{B}+\frac{2C_{,\theta}}{C}+
\tilde{G}\left(\frac{\dot{B}}{B}+\frac{2\dot{C}}{C}\right)\right\}\right]\\\label{B.6}&=&
\frac{1}{\lambda+2\xi R}\left\{8\pi q_{II}+\xi
L^{\nu}(\nabla_{\nu}\nabla_{\lambda}R)V^{\lambda}\right\}.
\end{eqnarray}
Contraction of Eq.(\ref{A.5}) with $K^{\mu}S^{\nu}$ and
$L^{\mu}S^{\nu}$ leads to
\begin{eqnarray}\nonumber
H_{1}&=&\frac{-1}{2B}\left[\Omega^{'}-\Omega\left\{\frac{C^{'}}{C}-
\frac{\tilde{G}\tilde{G}^{'}}{2(\sqrt{r^{2}B^{2}+\tilde{G}^{2}})}\right\}\right]
\frac{1}{6(\sqrt{r^{2}B^{2}+\tilde{G}^{2}})}\\\nonumber&\times&
\left[\left(2\sigma_{I}+\sigma_{II}\right)_{,\theta}+
\sigma_{I}\left\{\frac{B_{,\theta}}{B}+\frac{C_{,\theta}}{C}-
\tilde{G}\left(\frac{\dot{B}}{B}-\frac{\dot{C}}{C}\right)\right\}\right.\\\label{B.7}
&-&\left.
\sigma_{II}\left\{\frac{B_{,\theta}}{B}-\frac{2C_{,\theta}}{C}+
\tilde{G}\left(\frac{2\dot{B}}{B}-\frac{2\dot{C}}{C}\right)\right\}\right],
\\\nonumber
H_{2}&=&\frac{1}{6B}\left[\left(\sigma_{I}+2\sigma_{II}\right)^{'}+
\sigma_{I}\left\{\frac{2C^{'}}{C}-\frac{(rB)(rB)^{'}}
{r^{2}B^{2}+\tilde{G}^{2}}\right\}\right.\\\nonumber
&+&\left.\sigma_{II}\left\{\frac{C^{'}}{C}+\frac{2(rB)(rB)^{'}
+3\tilde{G}\tilde{G}^{'}}{2(r^{2}B^{2}+\tilde{G}^{2})}\right\}\right]
-\frac{1}{2\sqrt{r^{2}B^{2}+\tilde{G}^{2}}}\\\label{B.8}&\times&
\left[\Omega_{,\theta}-
\Omega\left\{\frac{C_{,\theta}}{C}+\tilde{G}\left(\frac{\dot{B}}{B}+
\frac{\dot{C}}{C}\right)\right\}\right].
\end{eqnarray}

Contraction of Eq.(\ref{A.7}) with $K^{\mu}K^{\nu},~
L^{\mu}L^{\nu},~ S^{\mu}S^{\nu}$ and $K^{\mu}L^{\nu}$ yield
\begin{eqnarray}\nonumber
&&\frac{1}{3}\dot{\varepsilon}_{I}+\frac{1}{9}\varepsilon_{I}\left(3\Theta+
\sigma_{II}-\sigma_{I}\right)
+\frac{1}{9}\varepsilon_{II}\left(2\sigma_{II}+\sigma_{I}\right)
-\Omega\varepsilon_{KL}-\frac{1}{\sqrt{(r^{2}B^{2}+\tilde{G}^{2})}}
\\\nonumber&\times&\left(H_{1,\theta}+H_{1}\frac{C_{,\theta}}{C}\right)
-\frac{H_{2}}{B}\left\{\frac{C^{'}}{C}-\frac{2(rB)(rB)^{'}
+\tilde{G}\tilde{G}^{'}}{2(r^{2}B^{2}+\tilde{G}^{2})}\right\}
\\\nonumber&=& -\frac{4\pi}{\lambda+2\xi R}\left\{\frac{1}{3}\rho \sigma_{I}
+\frac{1}{B}q_{I}^{'}+\frac{q_{II}}{\sqrt{r^{2}B^{2}+\tilde{G}^{2}}}
\left(\frac{\tilde{G}\dot{B}}{B}+\frac{B_{,\theta}}{B}\right)\right\}\\\nonumber
&+&\frac{\xi}{\lambda+2\xi
R}K^{\alpha}K^{\beta}\left\{\dot{R}\left(R_{\alpha\beta}-\frac{1}{3}R
g_{\alpha\beta}\right)-(\nabla_{\alpha}R)
R_{\beta\gamma}V^{\gamma}\right.\\\label{B.9}&+&\left.\nabla_{\alpha}
(\nabla_{\gamma}\nabla_{\beta}R)V^{\gamma}-
(\nabla_{\alpha}\nabla_{\beta}R)^{.}\right\}, \\\nonumber
&&\frac{1}{3}\dot{\varepsilon}_{II}+\frac{1}{9}\varepsilon_{II}\left(3\Theta+\sigma_{I}
-\sigma_{II}\right)
+\frac{1}{9}\varepsilon_{I}\left(2\sigma_{I}+\sigma_{II}\right)+\Omega\varepsilon_{KL}
+\frac{1}{B}\left(H_{2}^{'}+H_{2}\frac{C^{'}}{C}\right)\\\nonumber
&+&\frac{H_{1}}{\sqrt{(r^{2}B^{2}+\tilde{G}^{2})}}
\left\{\frac{C_{,\theta}}{C}-\frac{B_{,\theta}}{B}-\tilde{G}
\left(\frac{\dot{B}}{B}-\frac{\dot{C}}{C}\right)\right\}
=-\frac{4\pi}{\lambda+2\xi R}\\\nonumber
&\times&\left\{\frac{1}{3}\rho\sigma_{II}
+\frac{1}{2B}q_{I}\frac{(r^{2}B^{2}+\tilde{G}^{2})^{'}}{r^{2}B^{2}+\tilde{G}^{2}}
-\frac{\left(\tilde{G}\dot{q_{II}}+q_{II,\theta}\right)}{\sqrt{r^{2}B^{2}+\tilde{G}^{2}}}\right\}
\\\nonumber&+&\frac{\xi}{\lambda+2\xi
R}L^{\alpha}L^{\beta}\left\{\dot{R}\left(R_{\alpha\beta}-\frac{1}{3}R
g_{\alpha\beta}\right)-(\nabla_{\alpha}R)
R_{\beta\gamma}V^{\gamma}\right.\\\label{B.10}&+&\left.\nabla_{\alpha}
(\nabla_{\gamma}\nabla_{\beta}R)V^{\gamma}-
(\nabla_{\alpha}\nabla_{\beta}R)^{.}\right\}, \\\nonumber
&-&\frac{1}{3}{(\varepsilon_{I}+\varepsilon_{II})}^{.}-\frac{1}{3}\left(\varepsilon_{I}+\varepsilon_{II}\right)\Theta
-\frac{1}{9}\varepsilon_{I}\left(2\sigma_{II}+\sigma_{I}\right)
-\frac{1}{9}\varepsilon_{II}\left(2\sigma_{I}+\sigma_{II}\right)
\\\nonumber&+&\frac{1}{\sqrt{r^{2}B^{2}+\tilde{G}^{2}}}\left(H_{1,\theta}+H_{1}\frac{B_{,\theta}}{B}\right)
-\frac{1}{B}\left\{H_{2}^{'}+H_{2}\frac{(r^{2}B^{2}+\tilde{G}^{2})^{'}}{2(r^{2}B^{2}+\tilde{G}^{2})}\right\}
\\\nonumber&=&\frac{4\pi}{\lambda+2\xi R}
\left\{\frac{1}{3}\rho(\sigma_{I}+\sigma_{II})
-\frac{1}{B}q_{I}\frac{C^{'}}{C}-\frac{q_{II}}{\sqrt{r^{2}B^{2}+\tilde{G}^{2}}}
\left(\frac{\tilde{G}\dot{C}}{C}+\frac{C_{,\theta}}{C}\right)\right\}
\\\nonumber&+&\frac{\xi}{\lambda+2\xi
R}S^{\alpha}S^{\beta}\left\{\dot{R}\left(R_{\alpha\beta}-\frac{1}{3}R
g_{\alpha\beta}\right)-(\nabla_{\alpha}R)
R_{\beta\gamma}V^{\gamma}\right.\\\label{B.11}&+&\left.\nabla_{\alpha}
(\nabla_{\gamma}\nabla_{\beta}R)V^{\gamma}-
(\nabla_{\alpha}\nabla_{\beta}R)^{.}\right\}, \\\nonumber
&&2\dot{\varepsilon}_{KL}+{\varepsilon}_{KL}\left(2\Theta-\sigma_{I}-\sigma_{II}\right)+
\frac{\Omega}{3}\left(\varepsilon_{I}-\varepsilon_{II}\right)
+\frac{1}{B}\left[H_{1}'+H_{1}\right.\\\nonumber&\times&\left.\left\{\frac{2C'}{C}-
\frac{\left(2(Br)(Br)'+\tilde{G}\tilde{G}'\right)}{2(r^{2}B^{2}+\tilde{G}^{2})}\right\}\right]
-\frac{1}{\sqrt{r^{2}B^{2}+\tilde{G}^{2}}}\times\left[H_{2,\theta}+H_{2}\right.
\\\nonumber&\times&\left.
\left\{\frac{2C_{,\theta}}{C}-
\frac{2B_{,\theta}}{B}-\tilde{G}\left(\frac{\dot{B}}{B}-\frac{\dot{C}}{C}\right)\right\}\right]=
\frac{2\pi}{(\lambda+2\xi
R)\sqrt{r^{2}B^{2}+\tilde{G}^{2}}}\\\nonumber&\times&
\left\{q_{I}\left(\tilde{G}\frac{\dot{B}}{B}+\frac{B_{,\theta}}{B}\right)-
\tilde{G}\dot{q}_{I}-q_{I,\theta}\right\} +\frac{2\pi}{(\lambda+2\xi
R)B}\left\{-q_{II}'+q_{II}\frac{(r^{2}B^{2}+\tilde{G}^{2})'}{2(r^{2}B^{2}+\tilde{G}^{2})}\right\}
\\\nonumber&+&\frac{\xi}{\lambda+2\xi
R}K^{\alpha}L^{\beta}\left\{\dot{R}\left(R_{\alpha\beta}-\frac{1}{3}R
g_{\alpha\beta}\right)-(\nabla_{\alpha}R)
R_{\beta\gamma}V^{\gamma}\right.\\\label{B.12}&+&\left.\nabla_{\alpha}
(\nabla_{\gamma}\nabla_{\beta}R)V^{\gamma}-
(\nabla_{\alpha}\nabla_{\beta}R)^{.}\right\},
\end{eqnarray}
Addition of the above three equations gives
\begin{eqnarray}\nonumber
&&\frac{\tilde{G}}{\sqrt{r^{2}B^{2}+\tilde{G}^{2}}}
\left\{\frac{H_{2}\tilde{G}^{'}}{2B\sqrt{r^{2}B^{2}+\tilde{G}^{2}}}
+H_{1}\left(\frac{\dot{B}}{B}-\frac{\dot{C}}{C}\right)\right\}\\\nonumber
&=& \frac{4\pi}{B(\lambda+2\xi R)}
\left\{q_{I}^{'}+q_{I}\left(\frac{C^{'}}{C}+
\frac{(r^{2}B^{2}+\tilde{G}^{2})^{'}}{2(r^{2}B^{2}+\tilde{G}^{2})}\right)\right\}
\\\nonumber &+&\frac{4\pi}{\sqrt{r^{2}B^{2}+\tilde{G}^{2}}(\lambda+2\xi R)}
\left[q_{II}\left\{\tilde{G}\left(\frac{\dot{B}}{B}+\frac{\dot{C}}{C}\right)
+\frac{C_{,\theta}}{C}+\frac{B_{,\theta}}{B}\right\}+\tilde{G}\dot{q}_{II}+q_{II,\theta}\right]
\\\nonumber&+&\frac{\xi}{\lambda+2\xi
R}\left(K^{\alpha}K^{\beta}+L^{\alpha}L^{\beta}+S^{\alpha}S^{\beta}\right)
\left\{\dot{R}\left(R_{\alpha\beta}-\frac{1}{3}R
g_{\alpha\beta}\right)-(\nabla_{\alpha}R)
R_{\beta\gamma}V^{\gamma}\right.\\\label{B.13}&+&\left.\nabla_{\alpha}
(\nabla_{\gamma}\nabla_{\beta}R)V^{\gamma}-
(\nabla_{\alpha}\nabla_{\beta}R)^{.}\right\}.
\end{eqnarray}
The transport equation is
\begin{eqnarray}\label{B.14}
\tau
h^{\nu}_{\mu}q_{\nu;\alpha}^{tot}V^{\alpha}+q_{\mu}^{tot}={\kappa}
h^{\nu}_{\mu}T_{,\nu}+\frac{1}{2}{\kappa} T^{2}\left(\frac{\tau
V^{\alpha}}{{\kappa} T^{2}}\right)_{;\alpha}q_{\mu}^{tot}.
\end{eqnarray}

A combination of conservation equation and transport equation gives
\begin{eqnarray}\nonumber
&-&\tau\left(\frac{4}{3}\Theta
h_{\mu\nu}+\sigma_{\mu\nu}+\Omega_{\mu\nu}\right)q^{\nu}-\frac{\tau}{8\pi}
\left[\left\{\nabla^{\nu}\nabla_{\mu}(\lambda+2\xi R) -2\xi
R\right\}_{;\nu}\right.\\\nonumber&-&\left.\delta^{\nu}_{\mu}\left\{2\xi(\Box
R)_{;\nu}+\frac{1}{2}(\xi R^{2})_{;\nu}\right\}\right] + \tau
h^{\nu}_{\mu}q_{\nu;\alpha}^{(D)}V^{\alpha}+q_{\mu}^{(D)} +{\kappa}
h^{\nu}_{\mu}T_{,\nu}
\\\label{B.15}&+&\frac{1}{2}\kappa T^{2}\left(\frac{\tau
V^{\alpha}}{\kappa T^{2}}\right)_{;\alpha}q_{\mu}^{tot}.
\end{eqnarray}
Following are the terms used in the paper
\begin{eqnarray}\nonumber
D_{1}&=&\left[\frac{8\pi \rho q_{I}^{tot}}{3(\lambda+2\xi
R)}-\frac{\xi \nabla_{t}\nabla_{r}R}{B(\lambda+2\xi
R)}\right]\left[\frac{12\pi \rho}{(\lambda+2\xi
R)}-\frac{2R^{2}\xi}{(\lambda+2\xi R)}+\frac{\xi}{(\lambda+2\xi R)}
\right.\\\label{B.20}&\times&\left.\left\{2\left(g^{\delta\gamma}\nabla_{\delta}
\nabla_{\gamma}R-
V^{\delta}V^{\gamma}\nabla_{\delta}\nabla_{\gamma}R\right)+
\epsilon^{\nu\mu\gamma}\epsilon^{\delta}_{~\gamma\nu}
\nabla_{\delta}\nabla_{\mu}R\right\}\right]\\\nonumber
D_{2}&=&\left[\frac{8\pi \rho q_{II}^{tot}}{3(\lambda+2\xi
R)}-\frac{\xi
\nabla_{t}\nabla_{\theta}R}{\sqrt{\tilde{G}^{2}+r^{2}B^{2}}(\lambda+2\xi
R)}\right]\left[\frac{12\pi \rho}{(\lambda+2\xi
R)}-\frac{2R^{2}\xi}{(\lambda+2\xi
R)}\right.\\\nonumber&+&\left.\frac{\xi}{(\lambda+2\xi R)}
\left\{2\left(g^{\delta\gamma}\nabla_{\delta}\nabla_{\gamma}R-
V^{\delta}V^{\gamma}\nabla_{\delta}\nabla_{\gamma}R\right)+
\epsilon^{\nu\mu\gamma}\epsilon^{\delta}_{~\gamma\nu}
\nabla_{\delta}\nabla_{\mu}R\right\}\right]
\\\label{B.21}\\\nonumber
D_{3}&=&-\frac{1}{8\pi B}
\left[\left\{\nabla^{\nu}\nabla_{1}(\lambda+2\xi R) -2\xi R
R^{\nu}_{2}\right\}_{;\nu}-\delta^{\nu}_{1}\left\{2\xi(\Box
R)_{;\nu}+\frac{1}{2}(\xi R^{2})_{;\nu}\right\}\right]
\\\label{B.16}&+&\frac{T_{01}^{(D)}}{8\pi}\left\{\frac{1}{\tau}+\frac{1}{2}D_{t}\left[\ln
\left(\frac{\tau}{\kappa T^{2}}\right)\right]\right\}-\frac{1}{8\pi
B}\left(B T_{01}^{(D)}\right)_{;0}, \\\nonumber
D_{4}&=&-\frac{1}{8\pi \sqrt{r^{2}B^{2}+\tilde{G}^{2}}}
\left[\left\{\nabla^{\nu}\nabla_{2}(\lambda+2\xi R) -2\xi R
R^{\nu}_{2}\right\}_{;\nu}-\delta^{\nu}_{2}\left\{2\xi(\Box
R)_{;\nu}\right.\right.\\\nonumber&+&\left.\left.\frac{1}{2}(\xi
R^{2})_{;\nu}\right\}\right]
+\frac{(\tilde{G}T^{(D)}_{00}+T^{(D)})_{02}}{8\pi}\left\{\frac{1}{\tau}+
\frac{1}{2}D_{t}\left[\ln \left(\frac{\tau}{\kappa
T^{2}}\right)\right]\right\}\\\label{B.17}&-&\frac{1}{8\pi
\sqrt{r^{2}B^{2}+\tilde{G}^{2}}}\left(\tilde{G}T^{(D)}_{00}+T^{(D)}\right)_{;0}.
\\\label{B.18} D_{5}&=&\frac{\xi}{(\lambda+2\xi
R)}\left(K^{\alpha}K^{\beta}-\frac{1}{3}h_{\mu\nu}K^{\mu}K^{\nu}h^{\alpha\beta}\right)
\nabla_{\alpha}\nabla_{\beta}R,
\\\label{B.19} D_{6}&=&\frac{\xi}{(\lambda+2\xi
R)}\left(L^{\alpha}L^{\beta}-\frac{1}{3}h_{\mu\nu}L^{\mu}L^{\nu}h^{\alpha\beta}\right)
\nabla_{\alpha}\nabla_{\beta}R.
\\\label{B.19} D_{7}&=&\frac{\xi}{(\lambda+2\xi
R)}\left(L^{\alpha}L^{\beta}-\frac{1}{3}h_{\mu\nu}L^{\mu}L^{\nu}h^{\alpha\beta}\right)
\nabla_{\alpha}\nabla_{\beta}R,\\\nonumber
D_{8}&=&\frac{\xi}{(\lambda+2\xi
R)}K^{\alpha}K^{\beta}\left\{\dot{R}\left(R_{\alpha\beta}-\frac{1}{3}R
g_{\alpha\beta}\right)-(\nabla_{\alpha}R)
R_{\beta\gamma}V^{\gamma}\right.\\\label{B.23}&+&\left.\nabla_{\alpha}
(\nabla_{\gamma}\nabla_{\beta}R)V^{\gamma}-
(\nabla_{\alpha}\nabla_{\beta}R)^{.}\right\},\\\nonumber
D_{9}&=&\frac{\xi}{(\lambda+2\xi
R)}L^{\alpha}L^{\beta}\left\{\dot{R}\left(R_{\alpha\beta}-\frac{1}{3}R
g_{\alpha\beta}\right)-(\nabla_{\alpha}R)
R_{\beta\gamma}V^{\gamma}\right.\\\label{B.24}&+&\left.\nabla_{\alpha}
(\nabla_{\gamma}\nabla_{\beta}R)V^{\gamma}-
(\nabla_{\alpha}\nabla_{\beta}R)^{.}\right\}.
\end{eqnarray}

\end{document}